\documentclass{hcj}
\usepackage{mdwmath}
\usepackage{bbding}
\usepackage{pifont}
\usepackage{wasysym}
\usepackage{amssymb}
\usepackage{mdwtab}

\journalyear{2016}
\journalvolume{3}
\journalissue{1}
\journalpages{213-223}
\articledoi{10.15346/hc.v3i1.12}
\journalcopyright{}

\title{Conceptual Frameworks for Building  Online Citizen Science Projects}
\author{}
\author{Poonam Yadav\affil{Imperial College London}
       \and John Darlington\affil{Imperial College London}}
\authorrunning{P. Yadav, and J. Darlington}

\begin{document}

\maketitle

\begin{abstract}
In recent years, citizen science has grown in popularity due to a number of reasons, including the emphasis on informal learning and creativity potential associated with these initiatives. Citizen science projects address research questions from various domains, ranging from Ecology to Astronomy.  Due to the advancement of communication technologies, which makes outreach and engagement of wider communities easier, scientists are keen to turn their own research into citizen science projects. However, the development, deployment and management of these projects remains challenging. One of the most important challenges is building the project itself. There is no single tool or framework, which guides the step-by-step development of the project, since every project has specific characteristics, such as geographical constraints or volunteers' mode of participation. Therefore, in this article, we present a series of conceptual frameworks for categorisation, decision and deployment, which guide a citizen science project creator in every step of creating a new project starting from the research question to project deployment.  The frameworks are designed with consideration to the properties of already existing citizen science projects and could be easily extended to include other dimensions, which are not currently perceived.
\end{abstract}

\section{Introduction}
Citizen science projects encourage volunteers to participate in solving scientific problems across a wide range of scientific disciplines, from astronomy~\citep{DiskDetective} to ornithology~\citep{eBird, Nestwatch}.  The standard process in any scientific project involves three steps: scientific data collection, data processing and data-interpretation. Citizen science projects can involve volunteers (citizen scientists) in all three steps. These projects can be categorised based on the scientific applications' various attributes, e.g., user interaction, computational complexity and resource requirements~\citep{Wikipedia}. Citizen scientists are not necessarily knowledgeable about the particular scientific domain to which they contribute. Projects are often structured to enable participation by a wide range of individuals, from those who are scientific experts in the project's domain to members of the public who have no experience of the domain but are interested to learn and take part in the project. From this perspective, citizen science can present a range of educational benefits to the wider community in addition to helping to further scientific knowledge.

The participation of volunteers in citizen science project can be realised using different software-hardware configuration models (deployment models). The deployment scenarios describe the various configuration and set-up of citizen science applications for delivering services and aiding collaboration with volunteers. Among many citizen science applications' deployment categories, one important category is internet-based deployment.  An internet-based citizen science application (citizen cyberscience) deployment category represents a group of deployment scenarios where  volunteers can actively participate in science through interactive online interfaces (human computation) or passively by contributing computational power to projects (also known as volunteer computing). 

\begin{figure}[h]
 \centering
 \includegraphics[width=90mm,height=50mm]{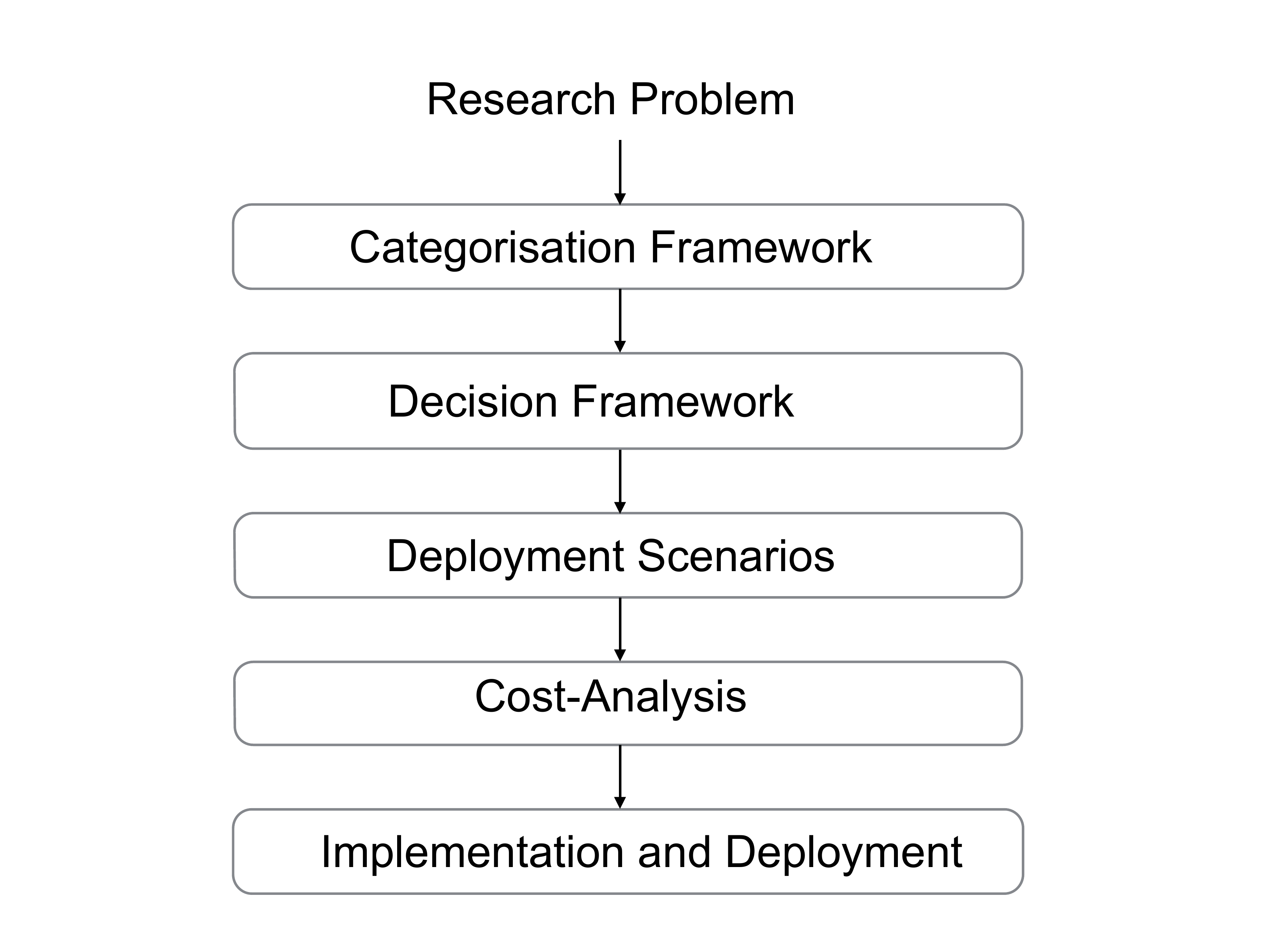} 
\caption{List of Conceptual Frameworks that are needed for  starting a Citizen Science Project}
\label{fig:ConceptualF} 
\end{figure}

However, due to diverse range of citizen science projects, the selection of a deployment and implementation models are not intuitive. In recent years, some domain-specific frameworks and tools have been designed~\citep{Tweddle2012, Shirk2012, Pocock2013, Pandya2012, Yadav2017CSGuidelines, Bonney2016}, however, frameworks that take into consideration the whole the citizen science projects landscape, are still missing~\citep{Bowser2014, Preece2016, Newman2012}. Therefore, in this article we analyse and present the conceptual frameworks (shown in Figure~\ref{fig:ConceptualF}), which help in guiding the deployment process of citizen science projects. To understand the deployment scenarios for citizen science applications, in this article  we identify the different attributes of these applications, which influence the selection of suitable computing platforms. We further categorise the applications based on the identified attributes and present connections between these different categories based on applications different attributes in a framework shown in Section~\ref{CategorisationFramework}.  In Section~\ref{DecisionFramework}, we present the decision framework which helps in  understanding and deciding whether the application can be deployed as a citizen science application. In Section~\ref{DeploymentFramework}, we present an association map that illustrates inter-links between attributes derived in Section~\ref{CategorisationFramework} and the computing platforms, which as a result presents the guiding routes (association links) for the possible deployment scenarios. In Section~\ref{CostAnalysis}, we discuss the cost analysis parameters that are involved in the successful deployment of a citizen science project.  In Section~\ref{Conclusions}, we present conclusion, scope and future work in this area.

\section{Categorisation Framework}\label{CategorisationFramework}
\begin{figure}[h]
 \centering
 \includegraphics[width=85mm,height=60mm]{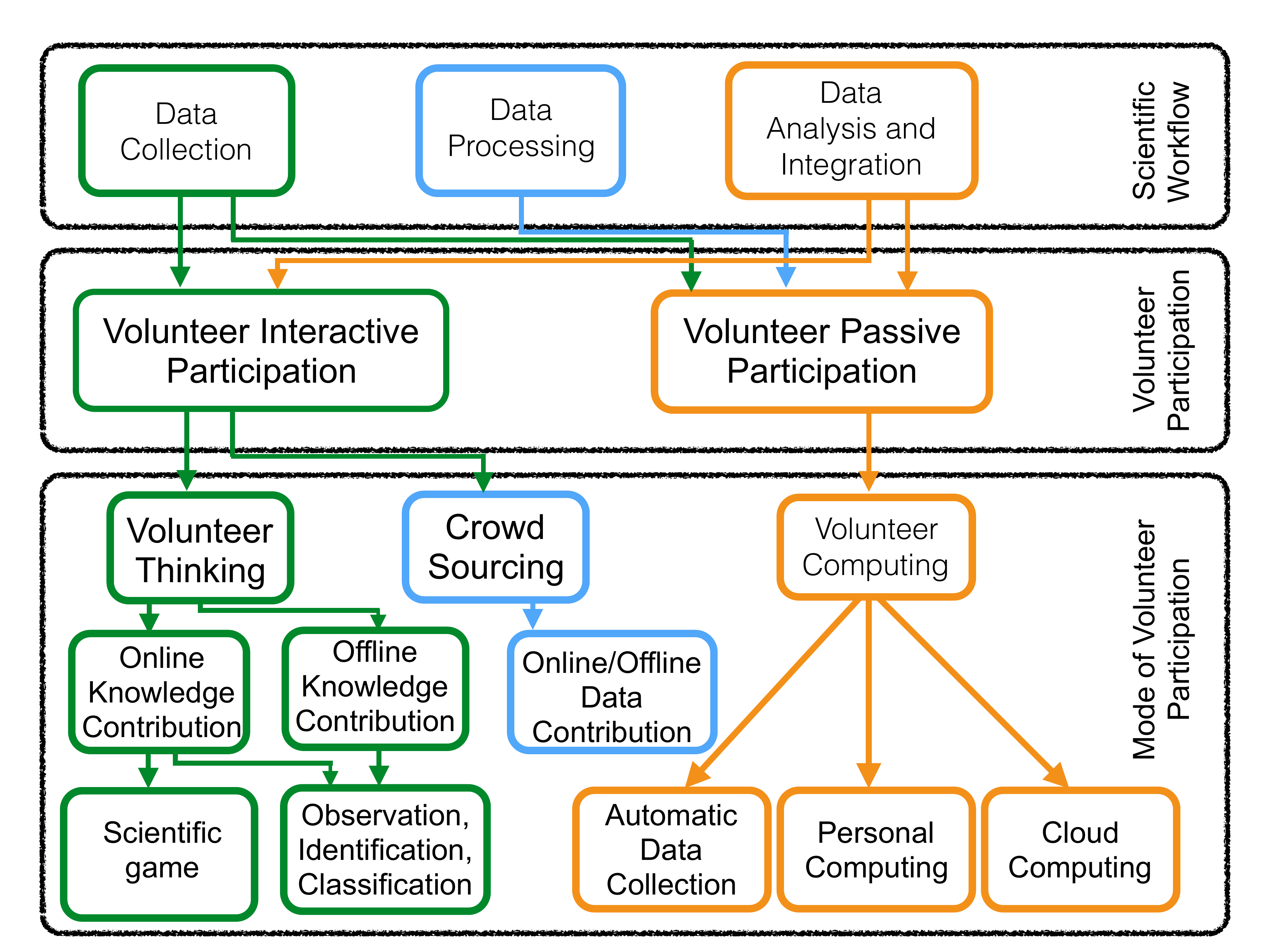} 
\caption{Citizen Science Applications' Categorisation Framework based on three categories: Scientific workflow, Volunteer participation, and Mode of Volunteer participation  (Computation and information processing).}
\label{fig:CF} 
\end{figure}

In this section, we present a categorisation framework (see Figure~\ref{fig:ConceptualF}).  Citizen science applications can be categorised by a range of different attributes such as the type of user interaction, scientific discipline, geographical location of the data and users, and whether the application is designed to capture/work with outdoor or indoor sources of data~\citep{Wikipedia,SciStarter2014}. However, not all attributes will necessarily contribute to determine the deployment scenario of a given application. We have, therefore, chosen only those attributes or features, which will help in defining and presenting a generalised set of application categories. This categorisation may be used as a helpful guide for application developers and research scientists who intend to exploit the benefits of human computation and volunteer computing. We list the following attributes, which we identified and will be used to categorise the applications: (a) Scientific workflow, (b) Volunteer participation, (c) Computation and information processing.
Based on these categories, the Figure~\ref{fig:CF} shows the categorisation framework, where the three attributes: scientific workflow, volunteer participation and information process represent the horizontal rows. The blocks within each attribute present different application categories. The links between the different blocks (categories) represent a route map or a logical association between the categories of different attributes. The logical association is useful for the citizen science application developers, which guide them in converting the scientific application to citizen science application.  We explain these attributes based categorisations in more detail in next sub-sections.

\subsection{Scientific Workflow}
Different scientific disciplines follow different scientific process workflows. These workflows can be highly complex and often become particularly challenging when they involve spatial or temporal variables such as location or time. To assist in the categorisation of workflow-based scientific applications for citizen science we have aimed to define a process structure. It consists of three general workflow phases in order to represent the vast majority of workflow-based citizen science applications.  These three process steps or workflow phases are: (a) Scientific data collection,  (b) Scientific data processing, (c) Scientific results (data) interpretation.
Each phase represents the application category based on the scientific workflow. A citizen science application involves one or two consecutive phases. For example, a citizen science application may require both data collection and data processing to be done by volunteers. In some scientific applications, it is difficult to distinguish data processing phase from data interpretation phase, therefore, the other information, e.g, the need of volunteer participation, decides which phase is more likely to be represented by the application.  However, from volunteer perspective this categorisation is irrelevant, but it is very useful for the citizen science application developer. The detail descriptions of the phases (categories) are given in the following sub-sections.

\subsubsection{Scientific Data Collection}  
Data collection is a critical task for many scientific projects, and yet is often time consuming and even tedious. It requires planning and preparation of data formats and data collection methods, correctly designed and configured data collection devices, and verification to ensure that required data precision and related quality parameters are met~\citep{Wiggins2016}. 

In some cases, complex experiments must be set up before the data can be collected, however, in many disciplines, initial feasibility studies also require data collection or surveys to be conducted before actual data is collected. Therefore, the data collection step may need to be repeated a number of times within a single project or may be an on-going task. Data may be in any number of different forms and some types of distributed data collection rely on human thinking to produce/identify the data. This is one particular area where citizen scientists, and the process of citizen science can excel. There are many situations where information is difficult, time consuming and expensive for a machine to extract but relatively easy for the human brain to identify. Such situations are often ideally suited to citizen science projects, taking advantage of the ability of the participating individuals to extract and provide data.
\subsubsection{	Scientific Data Processing}
Once scientists have collected the data, they create procedures and methods to process this data. In some projects, for example the SETI@Home~\citep{Seti} project where data is collected by a telescope, computationally intensive processing is required.
In projects such as this, citizen scientists contribute their computing power, a process known as volunteer computing, to help provide the vast amounts of computing power that are needed to process the captured data~\citep{Tinati2015}. 
\subsubsection{Scientific Results (Data) Interpretation}
Once data has been captured and processed, it must be interpreted. This is generally done through some form of visualisation or automated analysis. In some projects, data collection requires limited computational processing but a large amount of human cognition in order to interpret the meaning behind the outputs of the data processing stage. The projects such as GalaxyZoo~\citep{GalaxyZoo} require volunteers to identify galaxy structure in the images captured by the telescope. This task can be done by an automated computing program. The designing of an automated modelling-based approach to data interpretation is generally computationally intensive, however, these tasks can become affordable with the help of the volunteer scientists.
\subsection{Volunteer Participation}
Volunteers are invaluable resources in citizen science projects. These projects encourage volunteers to participate by contributing either their thinking power or their computing resources. The design and deployment scenarios for any citizen science application hugely depend on how volunteers are going to participate in the project. There is significant research work done around volunteer motivation, creativity, learning and volunteer continuous contribution~\citep{Charlene2013}. The citizen science projects that allow volunteers to be creative and learn need volunteers' active participation. Therefore, in this context, we split volunteer participation in two categories explained in the following sub-sections.
\subsubsection{Active Participation}
Internet and computer based active participation means that volunteers interactively contribute to a project by using their cognitive skills and their knowledge. It requires volunteers to dedicate time to actively engage with the task. For example, scientific projects hosted on the Zooniverse~\citep{Zooniverse}, CrowdCrafting~\citep{Crowdcrafting}  and Epicollect~\citep{Epicollect} platforms require the user to interact with an application via a web browser or a separate client interface. The linking of participation, platforms and deployment models is shown in  Figure~\ref{fig:CF} Figure~\ref{fig:DF} and, and is discussed in Section~\ref{DeploymentFramework}.
\subsubsection{Passive Participation}
Passive participation allows volunteers to contribute by providing and sharing computing resources through the Internet. In this type of participation, computationally intensive project tasks run on volunteers' computers, generally when they are otherwise idle, and with the permission of the volunteers. This doesn't require volunteers' time because they are not continuously interacting with the project. However, some volunteers do invest time into these project through interacting with the community, setting up a better system, etc., which is their combined (active and passive) participation. In order to identify idle periods, the tasks running on a volunteer' s computer monitor the system looking for periods of inactivity when volunteers are not interacting with their machine for their own work purposes.
\subsection{Participation Mode}
In this section, we group scientific applications into five categories based on the computation devices or resources they require, or the resources that volunteers contribute. First is volunteer cognition and thinking, where volunteer scientists only need to use their time, cognitive power and knowledge to analyse data. The volunteer participation in this category is marked as creative, learning oriented and interactive~\citep{Charlene2013}. To involve volunteers in this category, some projects present and structure scientific tasks as games that provide creative and learning-oriented participation, however, all citizen science games are not necessary to be creative. 

Similarly, the second category includes projects in which citizen scientists help in data collection by either filling in forms or using a device such as their mobile phone to collect data in the form of images, audio or other sensor data that may be accessible on the user's device. This data is later used by citizen science apps that may run on mobile devices. In the third category, volunteers install automatic data collection devices in their home, office or locations in which they are authorised to do so. These devices send data directly to the application server using any of the available wired or wireless data communication mechanisms. The fourth category involves volunteers downloading and running computationally intensive jobs on their personal computers when they are in idle mode. 

The last category is an interesting and rarely used participation mode where volunteers use their public or private cloud resources to run volunteer applications. This option frees the user from having the resources of their personal system consumed by a volunteer computing application but enables them to contribute to a project, often providing greater computational capacity since they are likely to provide a resource that is dedicated specifically to the volunteer computing application in question. 
\section{Decision Framework} \label{DecisionFramework}
The decision framework helps in understanding whether the application can be deployed as a citizen science application.  The decision framework for scientific data collection applications  has been well researched in the literature~\citep{Tweddle2012, Shirk2012, Pocock2013}. However, a decision framework which considers volunteer computing scenarios is not  researched. In this section, we present the decision framework which is a extended version of the strategic framework~\citep{Pocock2013} that includes volunteer  computing applications.  These are  a number of questions which helps in building the decision framework.
\begin{description}
\item{\textbf{Question 1 :}}  What are the spatial and temporal requirements of the project?\\ 
 A project is suitable for the citizen science approach if it requires volunteer participation from the wider geographical locations or volunteer contributions for a longer period of time. The projects which require volunteers from a specific areas  only for a shorter period of time, then citizen science is not an ideal approach. 
\item{\textbf{Question  2 :}}  What are the safety requirements of the project?\\  
A project is suitable for the citizen science approach if its tasks are safely carried by the volunteers~\citep{Bowser2014, Gellman}. 
\item{\textbf{Question 3 :}}   What is the required frequency of the participation? \\
If a project requires a number of tasks carried by the same volunteer, then the project is not suitable for the citizen science approach~\citep{Brasileiro2014,Cox2014, Jackson2015}.
\item{\textbf{Question 4 :}}  What is the project's task completion time? Can you parallelise the tasks?\\
The task size of a project is an important consideration. If  a task require volunteers to contribute a significant  amount of time (more than few minutes for a task) then it is not suitable for citizen science approach.  If a project requires only computing resources, it can be considered for a volunteer computing  (citizen science)  if it can be divided into parallel tasks~\citep{Pocock2013, Jackson2015, Nov2011}.
\end{description}
Once the citizen science approach is find suitable for a scientific project, then next step is to find the available computing resources to deploy the project as a citizen science project.  
\section{Deployment Framework}\label{DeploymentFramework}
\begin{figure}[h]
 \centering
 \includegraphics[width=85mm,height=60mm]{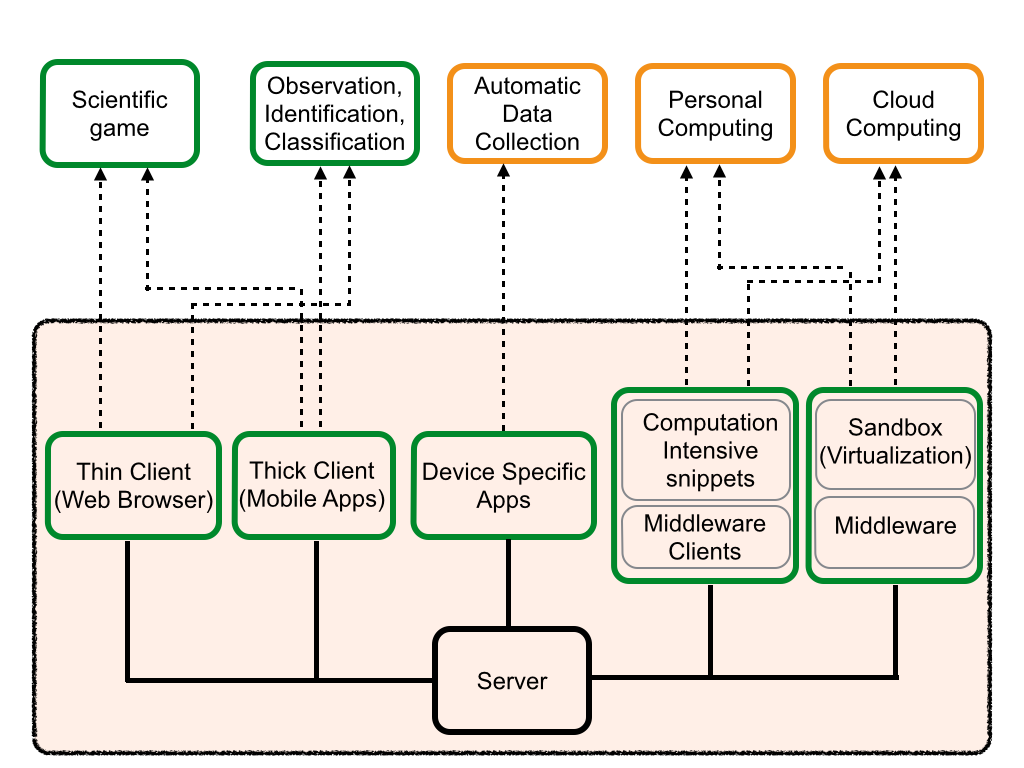} 
\caption{Citizen Science Applications' Deployment Framework}
\label{fig:DF} 
\end{figure}

The participation of volunteers in citizen science projects can be supported by a set of software-hardware configuration models (deployment models). These deployment models are represented by a group of corresponding scenarios that describe various configurations and methods for setting up of citizen science applications to deliver services~\citep{Yadav2014Deployment, Yadav2017CSdeployment, Yadav2017CSGuidelines, Parrish2015}. The deployment scenarios are presented using the links between the computation and information processing resources that were shown in the categorisation framework (see Figure~\ref{fig:CF}). The range of software delivery models is shown in Figure~\ref{fig:DF}. Citizen science applications make use of client-server distributing computing models. As shown in Figure~\ref{fig:DF}, we have selected five client-server deployment models that are suitable for different citizen science applications; we describe each of these in following sub-sections.
\subsection{Thin Client (Web browser-based applications)}
This model suits applications where there is generally a very limited requirement for client-side processing. In this deployment model, the project runs a web server that delivers pages to a client's web browser and volunteers interact with the web application through this browser-based interface (a thin client), which runs on their devices. The server-side element of the application may undertake extensive processing but this is not of significance to the end-user of the application. 

The project owners may host the application on their own web application servers or use middleware platforms, such as Zooniverse~\citep{Zooniverse} or Crowdcrafting~\citep{Crowdcrafting} to host their applications. From Figure 4, it can be seen that applications requiring volunteer cognition and volunteer observation are generally deployed using this model.

Middleware platforms~\citep{Crowdcrafting, Epicollect} are designed to host multiple citizen science web-applications, often providing a range of features to simplify and support the development and operation of these applications. For instance, they will often maintain multi-tenant user databases and project databases. Some middleware platforms such as Epicollect~\citep{Epicollect} and Crowdcrafting~\citep{Crowdcrafting} provide different application templates and tools that help scientists to design their web-based interface within a structure that has been designed specifically for citizen science applications. 

\subsection{Thick Client Deployment (Platform specific applications)}
For this deployment method, the project provides a client application that runs on users devices. Rather than a web-based application, this is likely to be a native application client that is designed specifically for one or more computer platforms~\citep{Paulos13, Epicollect}. This gives the application developer/operator improved capabilities in accessing and making use of user's devices when compared to a web-based application. Subject to the user granting permission for the application to do so, such applications are likely to access to a device's local storage, system libraries and hardware. The application owner either maintains a web server or use mobile app store, from which volunteers download the project's application client to their own devices, e.g. mobile phones, laptops or tablets. 

\subsection{Sensor Data Processing Deployment}
In this scenario, the device specific code is distributed online or pre-packaged into a dedicated, embedded device before deployment. Data is either sent back to the project server via the Internet as it is collected, or it is stored locally and sent in aggregated blocks, either when the data collection task finishes, or at intervals within the data collection task, perhaps through manual servicing of the device by project volunteers. In this deployment scenario, where the first step is to download and install the application on the sensor device. For devices that have direct Internet connectivity, this may be done from the device itself. For devices that don't have on-board Internet connectivity it may be done by connecting the device to another Internet connected computing device, for example a laptop. The application and any other software dependencies can be downloaded from project website directly or possibly from third-party servers that host the application and/or its dependencies. Once the installation is finished, citizen scientists use the sensor device(s) to monitor environmental properties of an indoor or outdoor environment. Some sensor devices, e.g. those that have wireless connectivity, transfer data directly to the project server or, in the case of short-range, low power wireless communication capabilities, to a gateway device that in-turn passes the data to the project server. Projects such as OpensourceBeehives~\citep{CSSP2014,SmartSensor2014} and DIY (do it yourself) electronics-based projects~\citep{PublicLab} are good examples of this deployment scenario.
\subsection{Computationally Intensive Platform Specific Applications}
In this deployment scenario, volunteers run an application either on their local resource(s)~\citep{Toth2007}, or on their rented public resources, e.g, AWS EC2~\citep{AWS}.  This scenario focuses on computation rather than user interactivity and can be run on resources that would otherwise be idle to make use of their computing power. By making use of many such user-provided resources, an application can gain access to vast amounts of computation at low cost. The data that the project needs to process is split into packages that can be distributed and independently processed by clients. Generally, volunteer application's executable code is run using a supported middleware client, e.g, Boinc~\citep{BOINC}. The middleware client connects directly to the project's application server and downloads the next available computing job/task/work-unit. The application code processes the task locally and then sends the resulting data back to the project server using middleware client~\citep{Yadav2017CGVas, Yadav2017CitizenGrid}. In this processing model, volunteers participate passively, as illustrated through the connected lines in the categorisation framework (see Figure~\ref{fig:CF}) and the deployment scenarios (see Figure~\ref{fig:DF}).

\section{Cost Analysis }\label{CostAnalysis}
Once an appropriate deployment model is finalised, the next step is to implement the model. But before that, the important step comes  regarding the project costing. There are a number of questions that  need to be answered before implementing a citizen science project.
\begin{description}
\item{\textbf{Question A:}} What are the available funding resources?
\item{\textbf{Question B:}} What are the open-source technologies are available for implementing the online project platforms?
\item{\textbf{Question C:}} What is the estimated cost for recruiting and maintaining the volunteers?
\item{\textbf{Question D:}} What is the estimated  project deployment costs?
\end{description}
The actual cost of the project depends on many other factors  such as project scale and time duration, which are some times unknown at the start of the project. In this section we presented only some intuitive questions and future work is required which provide more accurate funding estimations.  Once the project cost analysis is performed, the final step is the implementation and deployment of the project using the technologies that are relevant for  the selected deployment model.  

\section{Conclusions}\label{Conclusions}
In this article,  we presented a series of conceptual frameworks for categorisation, decision and deployment  that are derived by inferential analysis of already existing citizen science projects. These frameworks  guide a citizen science project creator in every step of creating a new project starting from the research question to project deployment.  The categorisation framework is flexible in the sense that it can accommodate new categories or new categorisation attributes easily to accommodate the new applications that will not fit in the current framework.The mapping of different attributes based categorisation framework to deployment models presents a helpful guidance tool for citizen science application developers. It helps scientists and developers by allowing them to choose an appropriate deployment model for their applications without spending time and resources in the detailed investigation about so many other attributes those do not contribute in the decision. Thus, the attribute-based categorisation and deployment association-map presented in this article speedup the deployment process of a citizen science application. These deployment frameworks can also be extended to provide additional fine-grain implementation details about the technology options for each of the deployment scenarios.  The decision and cost analysis frameworks are designed with consideration to the properties of already existing citizen science projects and could be easily extended to include other dimensions, which are not currently perceived.

\section{Acknowledgments}
We thank our colleagues, Dr Jeremy Cohen and Ms Christine Simpson for productive discussions that informed this article, as well as  Citizen Cyberlab project partners for sharing their valuable experiences in building citizen science tools and platforms.
This research was funded by the EU project Citizen Cyberlab (Grant No 317705). 

\bibliography{hcj-citizenscience}

\end{document}